\documentclass[aps,eqsecnum,preprint,floats,epsf,epsfig,nofootinbib]{revtex4}
\usepackage{graphicx}
\usepackage{epsfig}

\begin{document}
\def\be{\begin{eqnarray}}
\def\en{\end{eqnarray}}
\def\non{\nonumber}
\def\la{\langle}
\def\ra{\rangle}
\def\pp{{\prime\prime}}
\def\nc{N_c^{\rm eff}}
\def\vp{\varepsilon}
\def\hep{\hat{\varepsilon}}
\def\drho{\bar\rho}
\def\deta{\bar\eta}
\def\a{{\cal A}}
\def\B{{\cal B}}
\def\c{{\cal C}}
\def\d{{\cal D}}
\def\e{{\cal E}}
\def\p{{\cal P}}
\def\t{{\cal T}}
\def\B{{\cal B}}
\def\L{{\cal L}}
\def\P{{\cal P}}
\def\S{{\cal S}}
\def\T{{\cal T}}
\def\C{{\cal C}}
\def\A{{\cal A}}
\def\E{{\cal E}}
\def\V{{\cal V}}
\def\CP{{\it CP}~}
\def\CPP{{\it CP}}
\def\up{\uparrow}
\def\dw{\downarrow}
\def\vma{{_{V-A}}}
\def\vpa{{_{V+A}}}
\def\smp{{_{S-P}}}
\def\spp{{_{S+P}}}
\def\lrpartial{\buildrel\leftrightarrow\over\partial}
\def\J{{J/\psi}}
\def\3bar{{\bf \bar 3}}
\def\6bar{{\bf \bar 6}}
\def\10bar{{\bf \ov{10}}}
\def\ov{\overline}
\def\Lqcd{{\Lambda_{\rm QCD}}}
\def\pr{{Phys. Rev.}~}
\def\prl{{ Phys. Rev. Lett.}~}
\def\pl{{ Phys. Lett.}~}
\def\np{{ Nucl. Phys.}~}
\def\zp{{ Z. Phys.}~}
\def\lsim{ {\ \lower-1.2pt\vbox{\hbox{\rlap{$<$}\lower5pt\vbox{\hbox{$\sim$}
}}}\ } }
\def\gsim{ {\ \lower-1.2pt\vbox{\hbox{\rlap{$>$}\lower5pt\vbox{\hbox{$\sim$}
}}}\ } }

\font\el=cmbx10 scaled \magstep2{\obeylines\hfill BNL-HET-05/16}

\font\el=cmbx10 scaled \magstep2{\obeylines\hfill September, 2005}

\vskip 1.5 cm

\centerline{\large\bf $CP$-violating asymmetries in $B^0$ decays
to $K^+ K^- K_{S(L)}^0$ and $K_S^0 K_S^0 K_{S(L)}^0$}
\bigskip
\centerline{\bf Hai-Yang Cheng,$^1$ Chun-Khiang Chua$^1$ and
Amarjit Soni$^2$}
\medskip
\centerline{$^1$ Institute of Physics, Academia Sinica}
\centerline{Taipei, Taiwan 115, Republic of China}
\medskip

\medskip
\centerline{$^2$ Physics Department, Brookhaven National
Laboratory} \centerline{Upton, New York 11973}
\medskip

\bigskip
\bigskip
\centerline{\bf Abstract}
\bigskip
\small

Decay rates and time-dependent and direct \CP asymmetries in the
decays $B^0\to K^+K^-K_{S(L)}$ and $K_S K_S K_{S(L)}$ are studied.
Resonant and nonresonant contributions to the three-body decays
are carefully investigated. Nonresonant effects on 2-body and
3-body matrix elements are constrained by QCD counting rules. The
predicted branching ratios are consistent with the data within the
theoretical and experimental errors, though the theoretical
central values are somewhat smaller than the experimental ones.
Owing to the presence of {\it color-allowed} tree amplitudes in
$B^0\to K^+K^-K_{S(L)}$, this penguin-dominated mode may be
subject to a potentially significant tree pollution and the
deviation of the mixing-induced \CP asymmetry from that measured
in $B\to J/\psi K_S$, namely, $\Delta \sin
2\beta_{K^+K^-K_{S(L)}}\equiv \sin 2\beta_{K^+K^-K_{S(L)}}-\sin 2
\beta_{J/\psi K_S}$, can be as large as ${\cal O}(0.10)$. In
contrast, the $K_SK_SK_{S(L)}$ modes appear theoretically very
clean in our picture with negligible theoretical errors in
$\Delta\sin 2\beta_{K_SK_SK_{S(L)}}$. Direct \CP asymmetries in
$K^+K^-K_{S(L)}$ and $K_S K_S K_{S(L)}$ modes are found to be very
small.

\eject

\section{Introduction}

Considerable activity in search of possible New Physics beyond the
Standard Model has recently been devoted to the measurements of
time-dependent \CP asymmetries in neutral $B$ meson decays into
final \CP eigenstates defined by
 \be
 {\Gamma(\ov B(t)\to f)-\Gamma(B(t)\to f)\over
 \Gamma(\ov B(t)\to f)+\Gamma(B(t)\to
 f)}=\S_f\sin(\Delta mt)+\A_f\cos(\Delta mt),
 \en
where $\Delta m$ is the mass difference of the two neutral $B$
eigenstates, $S_f$ monitors mixing-induced \CP asymmetry and
$\A_f$ measures direct \CP violation (in the BaBar notation,
$\C_f=-\A_f$). The time-dependent {\it CP} asymmetries in the
$b\to sq\bar q$ penguin-induced two-body decays such as $B^0\to
(\phi,\omega,\pi^0,\eta',f_0)K_S$ measured by BaBar
\cite{BaBarSf1,BaBarSf2} and Belle
\cite{BelleSf1,BelleSf2,BelleSf3} show some indications of sizable
deviations from the expectation of the SM where \CP asymmetry in
all above-mentioned modes should be equal to $S_{J/\psi
K_S}=0.687\pm0.032$ \cite{HFAG} with a small deviation {\it at
most} ${\cal O}(0.1)$ \cite{LS,Browder}. Based on the framework of
QCD factorization \cite{BBNS}, the mixing-induced \CP violation
parameter $S_f$ in the seven 2-body modes
$(\phi,\omega,\rho^0,\eta',\eta,\pi^0,f_0)K_S$ has recently been
quantitatively studied in \cite{CCSsin2beta} and
\cite{Buchalla,Beneke}. It is found that the sign of $\Delta
S_f\equiv -\eta_fS_f-S_{J/\psi K_S}$ ($\eta_f$ being the \CP
eigenvalue of the final state $f$) at short distances is positive
except for the channel $\rho^0K_S$. After including final-state
rescattering effects, the central values of $\Delta S_f$ become
positive for all the modes under consideration, but they tend to
be rather small compared to the theoretical uncertainties involved
so that it is difficult to make reliable statements on the sign at
present \cite{CCSsin2beta}.

Time-dependent {\it CP} asymmetries in the $b\to sq\bar q$ induced
three-body decays $B^0\to K^+K^-K_S$ and $K_SK_SK_S$ have also
been measured by $B$ factories
\cite{BaBarSf2,BaBarKKK,BaBarKKKL,BaBar3Ks,BelleSf2,BelleSf3,BelleKKK}
(see
Table \ref{tab:Data}). Three-body modes such as these were first
discussed by Gershon and Hazumi \cite{Gershon}. While $K_SK_SK_S$
has fixed $CP$-parity, $K^+K^-K_S$ is an admixture of $CP$-even
and $CP$-odd components, rendering its \CP analysis more
complicated. By excluding the major $CP$-odd contribution from
$\phi K_S$, the 3-body $K^+K^-K_S$ final state is primarily
$CP$-even. A measurement of the \CPP-even fraction $f_+$ in the
$B^0\to K^+K^-K_S$ decay yields $f_+=0.89\pm0.08\pm0.06$ by BaBar
\cite{BaBarSf2} and $0.93\pm0.09\pm0.05$ by Belle \cite{BelleSf3},
while the \CPP-odd fraction in $K^+K^-K_L$ is measured to be
$f_-=0.92\pm0.33^{+0.13}_{-0.14}\pm0.10$ by BaBar
\cite{BaBarKKKL}. Hence, while $\eta_f=1$ for the $K_SK_SK_S$
mode, $\eta_f=2f_+-1$ for $K^+K^-K_S$ and $\eta_f=-(2f_--1)$ for
$K^+K^-K_L$. It is convenient to define an effective $\sin 2\beta$
via $S_f\equiv -\eta_f\sin 2\beta_{\rm eff}$. The results of $\sin
2\beta_{\rm eff}$ for $K^+K^-K_S$ obtained from the measurements
of $S_{K^+K^-K_S}$ and $f_+$ are also shown in Table
\ref{tab:Data}.

\begin{table}[t]
\caption{Mixing-induced \CP asymmetries $-S_f$ (top), direct \CP
violation $\A_f$ (middle) and branching ratios (in units of
$10^{-6}$, bottom) for $\ov B^0\to K^+K^-K_S$ and $K_SK_SK_S$
decays. For effective $\sin 2\beta$ for $K^+K^-K_S$, the third
error is due to the uncertainty in the fraction of \CPP-even
contributions to the decay rate. Experimental results are taken
from
\cite{BaBarSf2,BaBarKKK,BaBarKKKL,BaBar3Ks,BelleSf2,BelleSf3,BelleKKK}.
} \label{tab:Data}
\begin{ruledtabular}
\begin{tabular}{c r r r}
Final State & BaBar & Belle & Average  \\
\hline
 $K^+K^-K_S$\footnotemark[1] & $0.42\pm0.17\pm0.03$ &
 $0.52\pm0.16\pm0.03$ & $0.47\pm0.12$  \\
 $(\sin2\beta_{\rm eff})_{K^+K^-K_S}$ & $0.55\pm0.22\pm0.04\pm0.11$
 & $0.60\pm0.18\pm0.04^{+0.19}_{-0.12}$ &
 $0.57^{+0.18}_{-0.17}$ \\
 $K^+K^-K_L$\footnotemark[2] & $0.07\pm0.28^{+0.11}_{-0.12}$ & & $0.07\pm0.30$ \\
 $(\sin2\beta_{\rm eff})_{K^+K^-K_L}$ &
 $0.09\pm0.33^{+0.13}_{-0.14}\pm0.10$ & & $0.09\pm0.37$ \\
 $K_SK_SK_S$ & $0.63^{+0.28}_{-0.32}\pm0.04$ &
 $0.58\pm0.36\pm0.08$ & $0.61\pm0.23$ \\
 \hline
 $K^+K^-K_S$\footnotemark[1] & $-0.10\pm0.14\pm0.04$ & $-0.06\pm0.11\pm0.07$ &
 $-0.08\pm0.10$ \\
 $K^+K^-K_L$\footnotemark[2] & $-0.54\pm0.22^{+0.09}_{-0.08}$ & &
 $-0.54\pm0.24$ \\
 $K_SK_SK_S$ & $0.10\pm0.25\pm0.05$ &
 $0.50\pm0.23\pm0.06$ & $0.31\pm0.17$ \\
 \hline
 $K^+K^-K_{S}$ & $11.9\pm1.0\pm0.8$ & $14.2\pm1.7\pm2.0$ &
 $12.4\pm1.2$ \\
 $K_SK_SK_S$ & $6.9^{+0.9}_{-0.8}\pm0.6$ &
 $4.2^{+1.6}_{-1.3}\pm0.8$ & $6.2\pm1.2$~\footnotemark[3] \\
\end{tabular}
\end{ruledtabular}
\footnotetext[1]{with $\phi(1020)K_S$ excluded.}
\footnotetext[2]{with $\phi(1020)K_L$ excluded.}
\footnotetext[3]{with the error enlarged by a factor of $S=1.4$.}
\end{table}

In order to see if the current measurements of  the deviation of
$\sin 2\beta_{\rm eff}$ in $KKK$ modes from $\sin 2 \beta_{J/\psi
K_S}$ signal New Physics in $b\to s$ penguin-induced modes, it is
of great importance to examine and estimate how much of the
deviation of $\sin 2\beta_{\rm eff}$ is allowed in the SM. One of
the major uncertainties in the dynamic calculations lies in the
hadronic matrix elements which are nonperturbative in nature. One
way to circumvent this difficulty is to impose SU(3) flavor
symmetry \cite{Grossman03,Engelhard} or isospin and U-spin
symmetries \cite{Gronau} to constrain the relevant hadronic matrix
elements. While this approach is model independent in the symmetry
limit, deviations from that limit can only be computed in a model
dependent fashion.  In addition, it may have some weakness as
discussed in \cite{Engelhard}.

We shall apply the factorization approach in this work as it seems
to work even in the case of three-body $B$ decays~\cite{DKK}. By
using factorization and kaon time-like form factors extracted from
the $e^+e^-\to K\ov K$ process, the predicted $\ov B {}^0\to
D^{(*)+}K^- K^0$ rate agrees well with the data~\cite{DKK}. In
general, three-body $B$ decays are more complicated than the
two-body case as they receive resonant and nonresonant
contributions and involve 3-body matrix elements. Nonresonant
charmless three-body $B$ decays have been studied extensively
\cite{Deshpande,Fajfer1,Fajfer2,Deandrea1,Deandrea,Fajfer3} based
on heavy meson chiral perturbation theory (HMChPT)
\cite{Yan,Wise,Burdman}. However, the predicted decay rates are in
general unexpectedly large. For example, the branching ratio of
the nonresonant decay $B^-\to \pi^+\pi^-\pi^-$ is predicted to be
of order $10^{-5}$ in \cite{Deshpande} and \cite{Fajfer1}, which
is too large compared to the BaBar's preliminary result
$(0.68\pm0.41)\times 10^{-6}$ \cite{BaBar3pi}. The issue has to do
with the applicability of HMChPT. In order to apply this approach,
two of the final-state pseudoscalars have to be soft. The momentum
of the soft pseudoscalar should be smaller than the chiral
symmetry breaking scale $\Lambda_\chi\sim 830$ MeV. For 3-body
charmless $B$ decays, the available phase space where chiral
perturbation theory is applicable is only a small fraction of the
whole Dalitz plot. Therefore, it is not justified to apply chiral
and heavy quark symmetries to a certain kinematic region and then
generalize it to the region beyond its validity. In order to have
a reliable prediction for the  total rate of direct 3-body decays,
one should try to utilize chiral symmetry to a minimum. Therefore,
we will apply HMChPT only to the strong vertex and use the form
factors to describe the weak vertex \cite{Cheng:2002qu}. Moreover,
we shall introduce a form factor to take care of the off-shell
effect.

As shown in \cite{CCSsin2beta}, among the aforementioned seven
neutral $PK_S$ modes, only the $\omega K_S$ and $\rho^0 K_S$ modes
are expected to have a sizable deviation of the mixing-induced \CP
asymmetry $S_f$ from $S_{J/\psi K_S}$. More precisely, it is found
$\Delta S_{\omega K_S}=0.12^{+0.05}_{-0.06}$ and $\Delta
S_{\rho^0K_S}=-0.09^{+0.03}_{-0.07}$ \footnote{Note that since
$K^+K^-K_S$ is not a pure \CP eigenstate, we define $\Delta \sin
2\beta_{\rm eff}\equiv \sin 2 \beta_{\rm eff}-\sin 2\beta_{J/\psi
K}$ with $\sin 2\beta_{\rm eff}=-S_f/\eta_f$. In general, the
relation $\Delta S_f=\Delta \sin 2\beta_f^{\rm eff}$ holds for the
final state with fixed \CP parity.}
in the absence of final-state interactions \cite{CCSsin2beta}.
Although the tree contribution in these two modes is color
suppressed, the large cancellation between $a_4$ and $a_6$ penguin
terms renders the tree pollution relatively significant. Unlike
the above-mentioned case for two-body decays, the tree
contribution to the 3-body decay $B^0\to K^+K^-K_S$ is {\it
color-allowed} and hence it has the potential for producing a
large deviation from $\sin 2\beta$ measured in $B\to J/\psi K_S$.
We shall see in this work that it is indeed the case. In contrast,
the absence of tree pollution in $K_SK_SK_S$ renders it
theoretically very clean in our picture.

The layout of the present paper is as follows. In Sec. II we apply
the factorization approach to study $B^0\to K^+K^-K_S$ and
$K_SK_SK_S$ decays and discuss resonant and nonresonant
contributions in Sec. II. Numerical results for decay rates and
\CPP-violating parameters $S_f$ and $A_f$ and discussions are
presented in Sec. III. Sec. IV contains our conclusions.

\section{Formalism for charmless 3-body $B$ decays}

In the factorization approach, the matrix element of the $\ov
B\to\ov K\,\ov K K$ decay amplitude is given by
 \be \label{eq:factamp}
 \la \overline K\, \overline K K|{\cal H}_{\rm eff}|\ov B\ra
 =\frac{G_F}{\sqrt2}\sum_{p=u,c}\lambda_p \la\overline K\, \overline K K|T_p|\ov B\ra,
 \en
where $\lambda_p\equiv V_{pb} V^*_{ps}$ and \cite{BBNS}
 \be \label{eq:Tp}
 T_p&=&
 a_1 \delta_{pu} (\bar u b)_{V-A}\otimes(\bar s u)_{V-A}
 +a_2 \delta_{pu} (\bar s b)_{V-A}\otimes(\bar u u)_{V-A}
 +a_3(\bar s b)_{V-A}\otimes\sum_q(\bar q q)_{V-A}
 \non\\
 &&+a^p_4\sum_q(\bar q b)_{V-A}\otimes(\bar s q)_{V-A}
   +a_5(\bar s b)_{V-A}\otimes\sum_q(\bar q q)_{V+A}
       \non\\
 &&-2 a^p_6\sum_q(\bar q b)_{S-P}\otimes(\bar s q)_{S+P}
 +a_7(\bar s b)_{V-A}\otimes\sum_q\frac{3}{2} e_q (\bar q q)_{V+A}
 \non\\
 &&-2a^p_8\sum_q(\bar q b)_{S-P}\otimes\frac{3}{2} e_q
             (\bar s q)_{S+P}
 +a_9(\bar s b)_{V-A}\otimes\sum_q\frac{3}{2}e_q (\bar q q)_{V-A}\non\\
 &&+a^p_{10}\sum_q(\bar q b)_{V-A}\otimes\frac{3}{2}e_q(\bar s
 q)_{V-A},
 \en
with $(\bar q q')_{V\pm A}\equiv \bar q\gamma_\mu(1\pm\gamma_5)
q'$, $(\bar q q')_{S\pm P}\equiv\bar q(1\pm\gamma_5) q'$ and a
summation over $q=u,d,s$ being implied. The matrix element
$\la\overline K\, \overline K K|j\otimes j'|\ov B\ra$ corresponds
to $\la\overline K K|j|\ov B\ra\la\overline K|j'|0\ra$,
$\la\overline K|j|\ov B\ra \la\overline K K|j'|0\ra$ or
$\la0|j|\ov B\ra\la\overline K\, \overline K K|j'|0\ra$, as
appropriate, and $a_i$ are the NLO effective Wilson coefficients.
In this work, we take
 \be
 && a_1\approx0.99\pm0.37 i,\quad a_2\approx 0.19-0.11i, \quad a_3\approx -0.002+0.004i, \quad a_5\approx
 0.0054-0.005i,  \non \\
 && a_4^u\approx -0.03-0.02i, \quad a_4^c\approx
 -0.04-0.008i,\quad
 a_6^u\approx -0.06-0.02i, \quad a_6^c\approx -0.06-0.006i,
 \non\\
 &&a_7\approx 0.54\times 10^{-4} i,\quad a_8^u\approx (4.5-0.5i)\times
 10^{-4},\quad
 a_8^c\approx (4.4-0.3i)\times
 10^{-4},   \\
 && a_9\approx -0.010-0.0002i,\quad
 a_{10}^u \approx (-58.3+ 86.1 i)\times10^{-5},\quad
 a_{10}^c \approx (-60.3 + 88.8 i)\times10^{-5}, \non
 \en
for typical $a_i$ at the renormalization scale $\mu=m_b/2=2.1$~GeV
which we are working on.

Applying Eqs. (\ref{eq:factamp}), (\ref{eq:Tp}) and the equation
of motion, we obtain the $\ov B {}^0\to K^+ K^- \ov K {}^0$ decay
amplitude as
 \be
 \la\overline K {}^0 K^+ K^-|T_p|\ov B\ra&=&
 \la K^+\ov K {}^0|(\bar u b)_{V-A}|\ov B {}^0\ra \la K^-|(\bar s u)_{V-A}|0\ra
 \left[a_1 \delta_{pu}+a^p_4+a_{10}^p-(a^p_6+a^p_8) r_\chi\right]
 \non\\
 &&+\la \ov K {}^0|(\bar s b)_{V-A}|\ov B {}^0\ra
                   \la K^+ K^-|(\bar u u)_{V-A}|0\ra
    (a_2\delta_{pu}+a_3+a_5+a_7+a_9)
                   \non\\
 &&+\la \ov K {}^0|(\bar s b)_{V-A}|\ov B {}^0\ra
                   \la K^+ K^-|(\bar d d)_{V-A}|0\ra
    \bigg[a_3+a_5-\frac{1}{2}(a_7+a_9)\bigg]
    \non\\
 &&+\la \ov K {}^0|(\bar s b)_{V-A}|\ov B {}^0\ra
                   \la K^+ K^-|(\bar s s)_{V-A}|0\ra
    \bigg[a_3+a^p_4+a_5-\frac{1}{2}(a_7+a_9+a^p_{10})\bigg]
    \non\\
 &&+\la \ov K {}^0|\bar s b|\ov B {}^0\ra
       \la K^+ K^-|\bar s s|0\ra
       (-2 a^p_6+a^p_8)
       \non\\
  &&  +\la K^+ K^-\ov K {}^0|(\bar s d)_{V-A}|0\ra
     \la 0|(\bar d b)_{V-A}|\ov B {}^0\ra
       \bigg(a^p_4-\frac{1}{2} a^p_{10}\bigg)
       \non\\
 &&  + \la K^+ K^-\ov K {}^0|\bar s\gamma_5 d|0\ra
       \la 0|\bar d\gamma_5 b|\ov B {}^0\ra
       (-2a^p_6+a^p_8),
 \label{eq:AKKKs}
 \en
with $r_\chi=2 m_K^2/(m_b m_s)$. In the factorization terms, the
$K\ov K$ pair can be produced through a transition from the $\ov
B$ meson or can be created from vacuum through $V$ and $S$
operators. There exist two weak annihilation contributions, where
the $\ov B$ meson is annihilated and a final state with three
kaons is created.
Note that the OZI suppressed matrix element $\la K^+ K^-|(\bar d
d)_{V-A}|0\ra$ is included in the factorization amplitude since it
could be enhanced through the long-distance pole contributions via
the intermediate vector mesons such as $\rho^0$ and $\omega$.

To evaluate the above amplitude, we need to consider the $\ov B\to
K\ov K$, $0\to K\ov K$ and $0\to \ov K\,\ov K K$ matrix elements,
the so-called two-meson transition, two-meson and tree-meson
creation matrix elements in addition to the usual one-meson
transition and creation ones. The two-meson transition matrix
element $\la \ov K {}^0 K^+|(\bar u b)_{V-A}|\ov B {}^0\ra$ has
the general expression~\cite{LLW}
 \be
 \la \ov K {}^0 (p_1) K^+(p_2)|(\bar u b)_{V-A}|\ov B {}^0\ra
 &=&i r
 (p_B-p_1-p_2)_\mu+i\omega_+(p_2+p_1)_\mu+i\omega_-(p_2-p_1)_\mu
 \non\\
 &&+h\,\epsilon_{\mu\nu\alpha\beta}p_B^\nu (p_2+p_1)^\alpha
 (p_2-p_1)^\beta.
 \en
This leads to
 \be
 && \la K^-(p_3)|(\bar s
 u)_{V-A}|0\ra \la\ov K {}^0 (p_1) K^+(p_2)|(\bar u b)_{V-A}|\ov B {}^0\ra \non\\
 &&\quad=-\frac{f_K}{2}\left[2 m_3^2 r+(m_B^2-s_{12}-m_3^2) \omega_+
 +(s_{23}-s_{13}-m_2^2+m_1^2) \omega_-\right],
 \en
where $s_{ij}\equiv (p_i+p_j)^2$. A pole model calculation of the
$\ov B^0\to \ov K^0K^+$ transition matrix element amounts to
considering the strong interaction $\ov B {}^0\to \ov K {}^0 \ov
B_s^*$ followed by the weak transition $\ov B_s^*\to K^+$ and the
result is \cite{Cheng:2002qu}
 \be \label{eq:BKKpole}
 && \left[\la K^-(p_3)|(\bar s
 u)_{V-A}|0\ra \la\ov K {}^0 (p_1) K^+(p_2)|(\bar u b)_{V-A}|\ov B {}^0\ra\right]_{\rm pole}
 \non\\
 &&\quad=\frac{f_K}{f_\pi}\frac{g\sqrt{m_B
 m_{B_s^*}}}{s_{23}-m^2_{B_s^*}}
 F(s_{23},m_{B_s^*}) F_1^{B_sK}(m_3^2)
 \bigg[m_B+\frac{s_{23}}{m_B}-m_B\frac{m_B^2-s_{23}}{m_3^2}
 \bigg(1-\frac{F_0^{B_sK}(m_3^2)}{F_1^{B_sK}(m_3^2)}\bigg)\bigg]\non\\
 &&\qquad\times\bigg[m_1^2+m_3^2-s_{13}+\frac{(s_{23}-m_2^2+m_3^2)(m_B^2-s_{23}-m_1^2)}{2
 m_{B_s^*}^2}\bigg],
 \en
where $g$ is a heavy-flavor independent strong coupling which can
be extracted from the recent CLEO measurement of the $D^{*+}$
decay width, $g=0.59\pm0.01\pm0.07$ \cite{CLEOg}, and $F^{B_s
K}_{0,1}$ are the $B_s\to K$ weak transition from factors in the
standard convention \cite{BSW}. Since $B_s^*$ can be far from the
mass shell, it is necessary to introduce a form factor
$F(s_{23},m_{B_s^*})$ to take into account the off-shell effect of
the $B_s^*$ pole. Following \cite{CCS}, it is parameterized as
$F(s_{23},m_{B_s^*})=(\Lambda^2-m_{B_s^*}^2)/(\Lambda^2-s_{23})$
with the cut-off parameter $\Lambda$ chosen to be
$\Lambda=m_{B^*_s}+\Lambda_{\rm QCD}$.

It is worth making a digression for a moment. In principle, one
can apply HMChPT {\it twice} to evaluate the form factors
$r,~\omega_+$ and $\omega_-$ \cite{LLW}. However, this will lead
to too large decay rates in disagreement with experiment
\cite{Cheng:2002qu}. This is because the use of HMChPT is reliable
only in the kinematic region where $K^+$ and $\ov K^0$ are soft.
Therefore, the available phase space where chiral perturbation
theory is applicable is very limited. If the soft meson result is
assumed to be applicable to the whole Dalitz plot, the decay rate
will be greatly overestimated. Therefore, we employ the pole model
to evaluate the aforementioned form factors. We shall apply HMChPT
only {\it once} to the $\ov B {}^0K {}^0 B_s^*$ strong vertex and
introduce a form factor to take care of the momentum dependence of
the strong coupling.

The resonant pole contributions to the form factors $r$,
$\omega_\pm$ and $h$ can be worked out from
Eq.~(\ref{eq:BKKpole}). In principle, there are also nonresonant
contributions to these form factors. It turns out that the leading
nonresonant contribution can be determined as follows. We notice
that the same $\ov B\to K\ov K$ two-meson transition matrix
element also appears in the decay $B^-\to D^0 K^0 K^-$ under
factorization \cite{DKK}. The data favors a $1^-$ configuration in
the $K^0 K^-$ pair ~\cite{DKKdata}. The corresponding two-meson
transition matrix element is dominated by the $\omega_-$ term.
Following~\cite{DKK} we shall include a nonresonant contribution
to $\omega_-$ parametrized as
 \be
 \omega^{NR}_-=\kappa\, \frac{2p_B\cdot p_2}{s^2_{12}},
 \label{eq:omegaNR}
 \en
and employ the $B^-\to D^0 K^0 K^-$ data and apply isospin
symmetry to the $\ov B\to K\ov K$ matrix elements to determine the
unknown parameter $\kappa$. The denominator in the above
parametrization is inspired by the QCD counting rule which gives
rise to a $1/s_{12}^2$ asymptotic behavior,\footnote{As explained
in \cite{DKK}, at least two hard gluon exchanges are needed: one
creating the $s\bar s$ pair in $\ov K^0K^+$, the other kicking the
spectator to catch up with the energetic $s$ quark to form the $K$
meson. This gives rise to a $1/s_{12}^2$ asymptotic behavior.}
while the numerator $p_B\cdot p_2=m_B E_{K^+}$ is motivated by the
observation that $K^+$ contains an energetic $u$ quark coming from
the $b\to u$ transition.

The matrix elements involving 3-kaon creation are given
by~\cite{Cheng:2002qu}
 \be \label{eq:KKKme}
 &&\hspace{-0.5cm}\la \ov K {}^0(p_1) K^+(p_2) K^-(p_3)|(\bar s d)_{V-A}|0\ra\la
 0|(\bar d b)_{V-A}|\ov B {}^0\ra
 \approx  0, \\
 &&\hspace{-0.5cm}\la \ov K {}^0(p_1) K^+(p_2) K^-(p_3)|\bar s\gamma_5
 d|0\ra\la
 0|\bar d\gamma_5 b|\ov B {}^0\ra=v\frac{ f_B m_B^2}{f_\pi m_b}
 \left(1-\frac{s_{13}-m_1^2-m_3^2}{m_B^2-m_K^2}\right)F^{KKK}(m_B^2),
 \non
 \en
where
 \be \label{eq:v}
 v=\frac{m_{K^+}^2}{m_u+m_s}=\frac{m_K^2-m_\pi^2}{m_s-m_d},
 \en
characterizes the quark-order parameter $\la \bar q q\ra$ which
spontaneously breaks the chiral symmetry. Both relations in Eq.
(\ref{eq:KKKme}) are originally derived in the chiral limit
\cite{Cheng:2002qu} and hence the quark masses appearing in Eq.
(\ref{eq:v}) are referred to the scale $\sim$ 1 GeV . The first
relation reflects helicity suppression which is expected to be
even more effective for energetic kaons. For the second relation,
we introduce the form factor $F^{KKK}$ to extrapolate the chiral
result to the physical region. Following \cite{Cheng:2002qu} we
shall take $F^{KKK}(q^2)=1/[1-(q^2/\Lambda^2_\chi)]$ with
$\Lambda_\chi=0.83$~GeV being a chiral symmetry breaking scale.

We now turn to the 2-kaon creation matrix element which can be
expressed in terms of time-like kaon current form factors as
 \be \label{eq:weakff}
 \la K^+(p_{K^+}) K^-(p_{K^-})|\bar q\gamma_\mu q|0\ra
 &=& (p_{K^+}-p_{K^-})_\mu F^{K^+K^-}_q,
 \non\\
 \la K^0(p_{K^0}) \ov K^0(p_{\bar K^0})|\bar q\gamma_\mu q|0\ra
 &=& (p_{K^0}-p_{\bar K^0})_\mu F^{K^0\bar K^0}_q.
 \en
The weak vector form factors $F^{K^+K^-}_q$ and $F^{K^0\bar
K^0}_q$ can be related to the kaon electromagnetic (e.m.) form
factors $F^{K^+K^-}_{em}$ and $F^{K^0\bar K^0}_{em}$ for the
charged and neutral kaons, respectively. Phenomenologically, the
e.m. form factors receive resonant and nonresonant contributions
and can be expressed by
 \be \label{eq:emff}
 F^{K^+K^-}_{em}= F_\rho+F_\omega+F_\phi+F_{NR}, \qquad
 F^{K^0\bar K^0}_{em}= -F_\rho+F_\omega+F_\phi+F_{NR}'.
 \en
It follows from Eqs. (\ref{eq:weakff}) and (\ref{eq:emff}) that
 \be
 F^{K^+K^-}_u&=&F^{K^0\bar K^0}_d=F_\rho+3 F_\omega+\frac{1}{3}(3F_{NR}-F'_{NR}),
 \non\\
 F^{K^+K^-}_d&=&F^{K^0\bar K^0}_u=-F_\rho+3 F_\omega,
 \non\\
 F^{K^+K^-}_s&=&F^{K^0\bar K^0}_s=-3 F_\phi-\frac{1}{3}(3 F_{NR}+2F'_{NR}),
 \label{eq:FKKisospin}
 \en
where use of isospin symmetry has been made.

The resonant and nonresonant terms in Eq. (\ref{eq:emff}) can be
parametrized as
 \be
 F_{h}(s_{23})=\frac{c_h}{m^2_h-s_{23}-i m_h \Gamma_h},
 \qquad
 F^{(\prime)}_{NR}(s_{23})=\left(\frac{x^{(\prime)}_1}{s_{23}}
 +\frac{x^{(\prime)}_2}{s_{23}^2}\right)
 \left[\ln\left(\frac{s_{23}}{\tilde\Lambda^2}\right)\right]^{-1},
 \en
with $\tilde\Lambda\approx 0.3$ GeV. The expression for the
nonresonant form factor is motivated by the asymptotic constraint
from pQCD, namely, $F(t)\to (1/t)[\ln(t/\tilde \Lambda^2)]^{-1}$
in the large $t$ limit \cite{Brodsky}. The unknown parameters
$c_h$, $x_i$ and $x'_i$ are fitted from the kaon e.m. data, giving
the best fit values (in units of GeV$^2$ for $c_h$) ~\cite{DKK}:
\begin{equation}
\begin{array}{lll}
c_\rho=3c_\omega=c_\phi=0.363,
  & c_{\rho(1450)}=7.98\times 10^{-3},\ \
  & c_{\rho(1700)}=1.71\times10^{-3},\ \
\\
c_{\omega(1420)}=-7.64\times 10^{-2},
  & c_{\omega(1650)}=-0.116,
  & c_{\phi(1680)}=-2.0\times10^{-2},
\\
\end{array}
\label{eq:cj}
\end{equation}
and
\begin{eqnarray}
x_1=-3.26~{\rm GeV}^2, \qquad x_2=5.02~{\rm GeV}^4,
 \qquad x'_1=0.47~{\rm GeV}^2,
 \qquad x'_2=0.
\label{eq:xy}
\end{eqnarray}
Note that the form factors $F_{\rho,\omega,\phi}$ in
Eqs.~(\ref{eq:emff}) and (\ref{eq:FKKisospin}) include the
contributions from the vector mesons
$\rho(770),\,\rho(1450),\,\rho(1700)$,
$\omega(782),\,\omega(1420),\,\omega(1650),$ $\phi(1020)$ and
$\phi(1680)$.
It is interesting to note that (i) the fitted values of $c_{V}$
are very close to the vector meson dominance expression
$g_{_{V\gamma}} g_{VKK}$ for $V=\rho,\omega,\phi$~\cite{DM2,PDG},
where $g_{_{V\gamma}}$ is the e.m. coupling of the vector meson
defined by $\la
V|j_{em}|0\ra=g_{V\gamma}\epsilon^*_V$~\footnote{The vector meson
e.m. couplings are given by $g_{\phi\gamma}=e_s m_\phi f_\phi$,
$g_{\rho\gamma}=[(e_u-e_d)/\sqrt2] m_\rho f_\rho$ and
$g_{\omega\gamma}=[(e_u+e_d)/\sqrt2] m_\omega f_\omega$ where
$e_q$ is the quark's charge and $f_V$ is the vector decay
constant.} and $g_{VKK}$ is the $V\to KK$ strong coupling with,
$-g_{\phi K^+K^-}\simeq g_{\rho K^+K^-}/\sqrt2= g_{\omega
K^+K^-}/\sqrt2\simeq3.03$,
and (ii) the vector-meson pole contributions alone yield
$F^{K^+K^-}_{u,s}(0)\approx 1,-1$ and $F^{K^+K^-}_d(0)\approx 0$
as the charged kaon does not contain the valence $d$
quark.~\footnote{The sign convention is fixed by using $\la
M(q\bar q',p)\ov M(q\bar q',p')|\bar q\gamma_\mu q|0\ra=\la
M(q\bar q',p)|\bar q\gamma_\mu q|M(q\bar q',-p')\ra=(p-p')_\mu
|F^{MM}_q|$ in the case of a real $F^{MM}_q$.}
The matrix element in the decay amplitude relevant for our purpose
then has the expression
 \be
 \la \ov K {}^0(p_1)|(\bar s b)_{V-A}|\ov B {}^0\ra \la K^+(p_2)
 K^-(p_3)|(\bar q q)_{V-A}|0\ra
 =(s_{12}-s_{13}) F_1^{BK}(s_{23}) F^{K^+K^-}_q (s_{23}).
  \en

We also need to specify the 2-body matrix element $\la K^+
K^-|\bar s s|0\ra$ induced from the scalar density. It receives
resonant and non-resonant contributions:
 \be \label{eq:KKsff}
 \la K^+(p_2) K^-(p_3)|\bar s s|0\ra
 &\equiv& f^{K^+K^-}_s(s_{23})=\sum_{i}\frac{m_i \bar f_i g^{i\to KK}}{m_i^2-s_{23}-i
 m_i\Gamma_i}+f_s^{NR},
 \non\\
 f_s^{NR}&=&\frac{v}{3}(3 F_{NR}+2F'_{NR})+v\frac{\sigma}{s_{23}^2}
 \left[\ln\left(\frac{s_{23}}{\tilde\Lambda^2}\right)\right]^{-1},
 \en
where the scalar decay constant $\tilde f_i$ is defined in $\la
i|\bar s s|0\ra=m_i \bar f_i$, $g^{i\to KK}$ is the $i\to KK$
strong coupling, and the nonresonant terms are related to those in
$F_s^{K^+K^-}$ through the equation of motion.\footnote{The use of
equations of motion also leads to
 \be
 f_s^{K^+K^-}=-v F_s^{K^+K^-}.
 \en
Note that the pole contribution to $F_s^{K^+K^-}$ should be
dropped in the above relation as it applies only to nonresonant
contributions.}
The main scalar meson pole contributions are those that have
dominant $s\bar s$ content and large coupling to $K\ov K$. It is
found in \cite{ANS} that among the $f_0$ mesons, only $f_0(980)$
and $f_0(1530)$ have the largest couplings with the $K\ov K$ pair.
Note that $f_0(1530)$ is a very broad state with the width of
order 1 GeV \cite{ANS}. To proceed with the numerical
calculations, we use $g^{f_0(980)\to KK}=1.5$~GeV,
$g^{f_0(1530)\to KK}=3.18$~GeV, $\Gamma_{f_0(980)}=80$~MeV,
$\Gamma_{f_0(1530)}=1.160$~GeV~\cite{ANS}, $\bar
f_{f_0(980)}(\mu=m_b/2)\simeq 0.39$~GeV~\cite{Cheng:2005ye} and
$\bar f_{f_0(1530)}\simeq \bar f_{f_0(980)}$.
The sign of the resonant terms is fixed by $f_s^{K^+ K^-}(0)=v$
from a chiral perturbation theory calculation (see, for example,
\cite{Cheng:1988va}).
It should be stressed that although the nonresonant contributions
to $f_s^{KK}$ and $F_s^{KK}$ are related through the equation of
motion, the resonant ones are different and  not related {\it a
priori}. To apply the equation of motion, the form factors should
be away from the resonant region.
In the large $s_{23}$-region, the nonresonant contribution
dominated by the $1/s_{23}$ term is far away from the resonant
one. In contrast, the $1/s^2_{23}$ term dominates in the low
$s_{23}$-region where resonant contributions cannot be ignored.
The $1/s^2_{23}$ term in $F_s$ is not necessarily conveyed to
$f_S$ through the equation of motion.
Hence, the $1/s^2_{23}$ term in Eq.~(\ref{eq:KKsff}) is
undetermined and a new parameter $\sigma$, which is expected to be
of similar size as $x_2$, is assigned and will be determined later
by fitting to the data. The corresponding matrix element is now
given by
 \be
 \la \ov K {}^0(p_1)|\bar s b|\ov B {}^0\ra \la K^+(p_2) K^-(p_3)|\bar s s|0\ra
 =\frac{m_B^2-m_K^2}{m_b-m_s} F_0^{BK}(s_{23})
 f_s^{K^+K^-}(s_{23}).
 \en

Collecting all the relevant matrix elements evaluated above, we
are ready to compute the amplitude $A(\ov B {}^0\to K_{S(L)} K^+
K^-)=\pm A(\ov B {}^0\to \ov K {}^0 K^+ K^-)/\sqrt2$. Since under
\CPP-conjugation we have $K_S (\vec p_1)\to K_S(-\vec p_1)$,
$K^+(\vec p_2)\to K^-(-\vec p_2)$ and $K^-(\vec p_3)\to K^+(-\vec
p_3)$, the $\ov B {}^0\to K_S K^+ K^-$ amplitude can be decomposed
into  \CPP-odd and \CPP-even components
 \be
 &&A[\ov B {}^0\to K_S(p_1) K^+(p_2) K^-(p_3)]=A(s_{12},s_{13},s_{23})=A_{CP-}+A_{CP+},\non\\
 &&A_{CP\pm}=\frac{1}{2}
 [A(s_{12},s_{13},s_{23})\pm A(s_{13},s_{12},s_{23})].
 \en
Correspondingly, we have
 \be
 \Gamma&=&\Gamma_{CP+}+\Gamma_{CP-},\non\\
 \Gamma_{CP\pm}&=&\frac{1}{(2\pi)^3}\frac{1}{32 m_B^3}\int
 |A_{CP\pm}|^2ds_{12} ds_{13}=\frac{1}{(2\pi)^3}\frac{1}{32 m_B^3}\int
 |A_{CP\pm}|^2ds_{12} ds_{23}.
 \label{eq:Gamma}
 \en
The vanishing cross terms due to the interference between \CPP-odd
and \CPP-even components can be easily seen from the
(anti)symmetric properties of the amplitude and the integration
variables under the interchange of $s_{12}\leftrightarrow s_{13}$.
Similar relations hold for the conjugated $B^0$ decay rate
$\bar\Gamma$. 
The $CP$-even fraction $f_+$ is defined by
 \be \label{eq:f+}
 f_+\equiv \left.{\Gamma_{CP+}+\ov \Gamma_{CP+} \over
 \Gamma+\ov\Gamma}\right|_{\phi K_S~{\rm excluded}.}
 \en
Note that results for the $K^+K^-K_L$ mode are identical to the
$K^+K^-K_S$ ones with the \CP eigenstates interchanged. For
example, results for $(K^+K^-K_L)_{CP+}$ are the same as those for
$(K^+K^-K_S)_{CP-}$ and hence $f_+$ in $K^+K^-K_S$ corresponds to
$f_-$ in $K^+K^-K_L$.

We next turn to the $\ov B {}^0\to K_S K_S K_S,\,K_S K_S K_L$
decays. The decay amplitudes are given by
 \be
 A[\ov B{}^0\to K_S(p_1) K_S(p_2) K_{S,L}(p_3)]
 &=&\left(\frac{1}{2}\right)^{3/2}
 \bigg\{\pm A[\ov B{}^0\to K^0(p_1) \ov K
 {}^0(p_2) \ov K {}^0(p_3)]
 \non\\
 &&\qquad\pm A[\ov B{}^0\to K^0(p_2) \ov K
 {}^0(p_3) \ov K {}^0(p_1)]
 \non\\
 &&\qquad+A[\ov B{}^0\to K^0(p_3) \ov K
 {}^0(p_1) \ov K {}^0(p_2)]\bigg\},
 \label{eq:AKsKsKs}
 \en
with
 \be
 A[\ov B{}^0\to K^0(p_1) \ov K
 {}^0(p_2) \ov K {}^0(p_3)]&=&\frac{G_F}{\sqrt2}\sum_{p=u,c}\lambda_p
 \Bigg\{
 \Big[\la K^0(p_1)\ov K {}^0(p_2)|(\bar d b)_{V-A}|\ov B {}^0\ra \la \ov K {}^0(p_3)|(\bar s d)_{V-A}|0\ra
 \non\\
 &&+\la K^0(p_1)\ov K {}^0(p_3)|(\bar d b)_{V-A}|\ov B {}^0\ra \la \ov K {}^0(p_2)|(\bar s
 d)_{V-A}|0\ra \Big]
 \non\\
 &&\times\Big(a^p_4+\frac{1}{2}a^p_{10}-(a^p_6-\frac{1}{2}a^p_8)
 r_\chi\Big)
 \non\\
 &&+\Big[\la \ov K {}^0 (p_2)|\bar s b|\ov B {}^0\ra
       \la K^0(p_1) \ov K {}^0(p_3)|\bar s s|0\ra
 \non\\
 &&    +\la \ov K {}^0 (p_3)|\bar s b|\ov B {}^0\ra
       \la K^0(p_1) \ov K {}^0(p_2)|\bar s s|0\ra\Big]
       (-2 a^p_6+a^p_8)
       \non\\
 &&  +   \la K^0(p_1) \ov K {}^0(p_2) \ov K {}^0(p_3)|\bar s\gamma_5 d|0\ra
      \la 0|\bar d\gamma_5 b|\ov B {}^0\ra
       (-2a^p_6+a^p_8)
 \non\\
       &&+\Big[\la \ov K {}^0 (p_2)|(\bar s b)_{V-A}|\ov B {}^0\ra
       \la K^0(p_1) \ov K {}^0(p_3)|(\bar s s)_{V-A}|0\ra
 \non\\
 &&    +\la \ov K {}^0 (p_3)|(\bar s b)_{V-A}|\ov B {}^0\ra
       \la K^0(p_1) \ov K {}^0(p_2)|(\bar s s)_{V-A}|0\ra\Big]
 \non\\
      &&\times \left[a_3+a^p_4+a_5-\frac{1}{2}(a_7+a_9+a_{10})\right]\Bigg\},
 \en
where the last term will not contribute to the purely \CPP-even
decay $\overline B {}^0\to K_S K_S K_S$. Decay rates for the $K_S
K_S K_S$ and $K_S K_S K_L$ modes can be obtained from
Eq.~(\ref{eq:Gamma}) with an additional factor of $1/3!$ and
$1/2!$, respectively, for identical particles in the final state.

We now consider the \CP asymmetries for $\ov B {}^0\to K^+ K^-
K_{S(L)},\,K_S K_S K_{S(L)}$ decays. The direct \CP asymmetry and
the mixing induced \CP violation are defined by
 \be \label{eq:A&S}
 \A_{KKK}&=& \left.\frac{\Gamma-\ov
 \Gamma}{\Gamma+\ov \Gamma}   \right.
 \non\\
       &=&\left. \frac{\int
 |A|^2ds_{12} ds_{23}-\int
 |\bar A|^2ds_{12} ds_{23}}{\int
 | A|^2ds_{12} ds_{23}+\int
 |\bar A|^2ds_{12} ds_{23}},\right.
 \non\\
 \S_{KKK,CP\pm}&=&   \left. \frac{2\int
 {\rm Im}(e^{-2i\beta} A_{CP\pm} \bar A^*_{CP\pm}) ds_{12} ds_{23}}{\int
 |A_{CP\pm}|^2ds_{12} ds_{23}+\int
 |\bar A_{CP\pm}|^2ds_{12} ds_{23}},\right.
 \non\\
 \S_{KKK}&=&  \left. \frac{2\int
 {\rm Im}(e^{-2i\beta} A \bar A^*) ds_{12} ds_{23}}{\int
 |A|^2ds_{12} ds_{23}+\int
 |\bar A|^2ds_{12} ds_{23}}\right. 
 \non\\
 &=&f_+\,S_{KKK,CP+}+(1-f_+)\,S_{KKK,CP-},
 \en
where $\bar A$ is the decay amplitude of $B^0\to K^+K^-K_{S(L)}$
or $K_SK_SK_{S(L)}$. For the $K^+K^-K_S$ mode, it is understood
that the contribution from $\phi K_S$ is excluded. It is expected
in the SM that $\S_{KKK,CP+}\equiv \sin 2 \beta_{\rm
eff}\approx\sin 2\beta$, $\S_{KKK,CP-}\approx -\sin 2\beta$ and
hence $\S_{KKK}\approx -(2f_+-1)\sin 2\beta$.\footnote{Writing the
\CPP-conjugated decay amplitude as $\bar A=\bar A_{CP+}+\bar
A_{CP-}$, we have $\bar A_{CP\pm}=\pm A_{CP\pm}$ with
$\lambda_p\to\lambda^*_p$. This leads to $\S_{KKK,CP-}\approx
-\S_{KKK,CP+}$.}

\section{Numerical results and discussions}

\begin{table}[t]
\caption{Branching ratios for $\ov B {}^0\to K^+ K^- K_S,\,K_S K_S
K_S,\,K_S K_S K_L$ decays and the fraction of \CPP-even
contribution to $\ov B^0\to K^+K^-K_S$, $f_+$ [see Eq.
(\ref{eq:f+})]. The branching ratio of \CPP-odd $K^+K^-K_S$ with
$\phi K_S$ excluded is shown in parentheses. Results for
$(K^+K^-K_L)_{CP\pm}$ are identical to those for
$(K^+K^-K_S)_{CP\mp}$. Theoretical errors correspond to the
uncertainties in (i) $\kappa$, (ii) $m_s$, $F^{BK}_0$ and $\sigma$
(constrained by the $K_SK_SK_S$ rate), and (iii) $\gamma$. }
\label{tab:Br}
\begin{ruledtabular}
\begin{tabular}{c r r}
Final State &${\cal B}(10^{-6})_{\rm theory}$ &${\cal B}(10^{-6})_{\rm expt}$ \\
\hline
 $K^+ K^- K_S$
       & $7.33^{+8.38+2.31+0.70}_{-1.08-1.59-0.10}$
       & $12.4\pm1.2$ \\
 $(K^+ K^- K_S)_{CP+}$
       & $5.45^{+5.29+1.48+0.05}_{-0.65-1.13-0.06}$
       &  \\
 $(K^+ K^- K_S)_{CP-}$
       & $1.88^{+3.08+0.83+0.04}_{-0.43-0.46-0.04}$
       &  \\
       & $(0.48^{+2.98+0.54+0.03}_{-0.40-0.22-0.03})$
       &  \\
 $K_S K_S K_S$
       & input
       & $6.2\pm1.2$ \\
 $K_S K_S K_L$
       & $5.74^{+6.02+2.24+0.02}_{-0.88-1.40-0.03}$
       &
       \\
 \hline
  &$f_+^{\rm theory}$  &$f_+^{\rm expt}$ \\
 \hline
 $K^+K^-K_S$
            & $0.92^{+0.06+0.04+0.00}_{-0.16-0.08-0.00}$
            & $0.91\pm0.07$
            \\
 \hline
 &$f_-^{\rm theory}$  &$f_-^{\rm expt}$ \\
 \hline
 $K^+K^-K_L$
            & $0.92^{+0.06+0.04+0.00}_{-0.16-0.08-0.00}$
            & $0.92\pm0.37$
 \end{tabular}
\end{ruledtabular}
\end{table}

\begin{table}[t]
\caption{Mixing-induced and direct \CP asymmetries $\sin
2\beta_{\rm eff}$ (top) and $\A_f$ (in $\%$, bottom),
respectively, in $B^0\to K^+K^-K_S$ and $K_SK_SK_S$ decays.
Results for $(K^+K^-K_L)_{CP\pm}$ are identical to those for
$(K^+K^-K_S)_{CP\mp}$. Experimental results are taken from Table
I. } \label{tab:AS}
\begin{ruledtabular}
\begin{tabular}{l r r}
 Final State & $\sin 2\beta_{\rm eff}$  & Expt.  \\
 \hline
 $(K^+K^-K_S)_{\phi K_S~{\rm excluded}}$
            & $0.749^{+0.080+0.024+0.004}_{-0.013-0.011-0.015}$
            & $0.57^{+0.18}_{-0.17}$
            \\
 $(K^+K^-K_S)_{CP+}$
            & $0.770^{+0.113+0.040+0.002}_{-0.031-0.023-0.013}$
            &
            \\
 $(K^+K^-K_L)_{\phi K_L~{\rm excluded}}$
            & $0.749^{+0.080+0.024+0.004}_{-0.013-0.011-0.015}$
            & $0.09\pm0.34$
            \\
 $K_SK_SK_S$
            & $0.748^{+0.000+0.000+0.007}_{-0.000-0.000-0.018}$
            & $0.65\pm0.25$
            \\
 $K_SK_SK_L$
            & $0.748^{+0.001+0.000+0.007}_{-0.001-0.000-0.018}$
            &
            \\
 \hline
  &$\A_f(\%)$  &Expt. \\
 \hline
 $(K^+K^-K_S)_{\phi K_S~{\rm excluded}}$
            & $0.16^{+0.95+0.29+0.01}_{-0.11-0.32-0.02}$
            & $-8\pm10$
            \\
 $(K^+K^-K_S)_{CP+}$
            & $-0.09^{+0.73+0.16+0.01}_{-0.00-0.27-0.01}$
            &
            \\
 $(K^+K^-K_L)_{\phi K_L~{\rm excluded}}$
            & $0.16^{+0.95+0.29+0.01}_{-0.11-0.32-0.02}$
            & $-54\pm24$
            \\
 $K_SK_SK_S$
            & $0.74^{+0.02+0.00+0.05}_{-0.06-0.01-0.06}$
            & $31\pm17$
            \\
 $K_SK_SK_L$
            & $0.77^{+0.12+0.08+0.06}_{-0.28-0.11-0.07}$
            &
            \\
 \end{tabular}
\end{ruledtabular}
\end{table}

To proceed with the numerical calculations, we need to specify the
input parameters. For the CKM matrix elements, we use the
Wolfenstein parameters $A=0.825$, $\lambda=0.22622$, $\bar
\rho=0.207$ and $\bar \eta=0.340$, corresponding to
$(\sin2\beta)_{CKM}=0.724$~\cite{CKMfitter}. For $B\to K$ form
factors we shall use those derived in the covariant light-front
quark model~\cite{CCH} with the assigned error to be $0.03$,
namely, $F_{0,1}^{BK}(0)=0.35\pm0.03$. The parameter $\kappa$ in
Eq.~(\ref{eq:omegaNR}) is determined from the $B^-\to D^0 K^0 K^-$
data. From the measured branching ratio ${\cal B}(B^-\to D^0 K^0
K^-)=(5.5\pm1.4\pm0.8)\times 10^{-4}$ \cite{DKKdata}, we obtain
$\kappa=3.1^{+5.1}_{-1.8}$~GeV where use of $a^{DKK}_1=0.935$ and
$a^{DKK}_2(\simeq a^{D\rho}_2)=0.4\pm 0.2$ has been
made~\cite{DKK}. For the quark masses and the unitarity angle
$\gamma$, we shall use $m_b(m_b)=4.2$ GeV,
$m_s(m_b/2)=80\pm20$~MeV and
$\gamma=(58.6\pm7)^\circ$~\cite{CKMfitter}. The $K_SK_SK_S$ rate
sensitive to the parameter $\sigma$ in Eq.~(\ref{eq:KKsff}) is
used to determine $\sigma=(-10.4^{+5.4}_{-4.8})$GeV$^4$, where the
errors include the uncertainties in the $K_SK_SK_S$ decay rate,
the strange quark mass and the $F_0^{BK}$ form factor.

Results for the decay rates and \CP asymmetries in $\ov B {}^0\to
K^+ K^- K_{S(L)},\,K_S K_S K_{S(L)}$ are exhibited in
Table~\ref{tab:Br} and Table~\ref{tab:AS}, respectively. The
theoretical errors shown there are from the uncertainties in (i)
the parameter $\kappa$ which governs the nonresonant contribution
to the form factor $\omega_-$ [see Eq. (\ref{eq:omegaNR})], (ii)
the strange quark mass $m_s$, the form factor $F^{BK}_0$ and
$\sigma$ [see Eq. (\ref{eq:KKsff})] constrained from the
$K_SK_SK_S$ rate, and (iii) the unitarity angle $\gamma$.
To compute the \CPP-even fraction $f_+$ and $\sin 2\beta_{\rm
eff}$ for $K^+K^-K_S$, we need to turn off the coefficient
$c_\phi$ in Eq.~(\ref{eq:FKKisospin}).  As one can see from Table
\ref{tab:Br}, the predicted rates for $\ov B {}^0\to K^+ K^-
K_{S(L)}$ decays and the $CP$-even (odd) ratio $f_{+(-)}$ are in
accordance with the data within errors, though the theoretical
central values on rates are somewhat smaller than the experimental
ones. Theoretical errors on the branching ratios are dominated by
the sizable error in $\kappa$ and the uncertainty in the strange
quark mass as the penguin term $a_6 r_\chi$ and the parameter $v$
are very sensitive to $m_s$. Note that the second error in rates
(including the contribution from the uncertainty in $\sigma$) are
constrained from the $K_SK_SK_S$ rate and hence are reduced
significantly. For the first error, we note that the larger the
value of $|\kappa|$ we have, the larger rate on \CPP-odd
$K^+K^-K_S$ is obtained, leading to a smaller value of
$f_+(K^+K^-K_S)$. Since the central value of our $f_+(K^+K^-K_S)$
agrees well with data, $\kappa$ is preferred to be around its
central value.

\begin{figure}[t]
  \centerline{\epsfig{figure=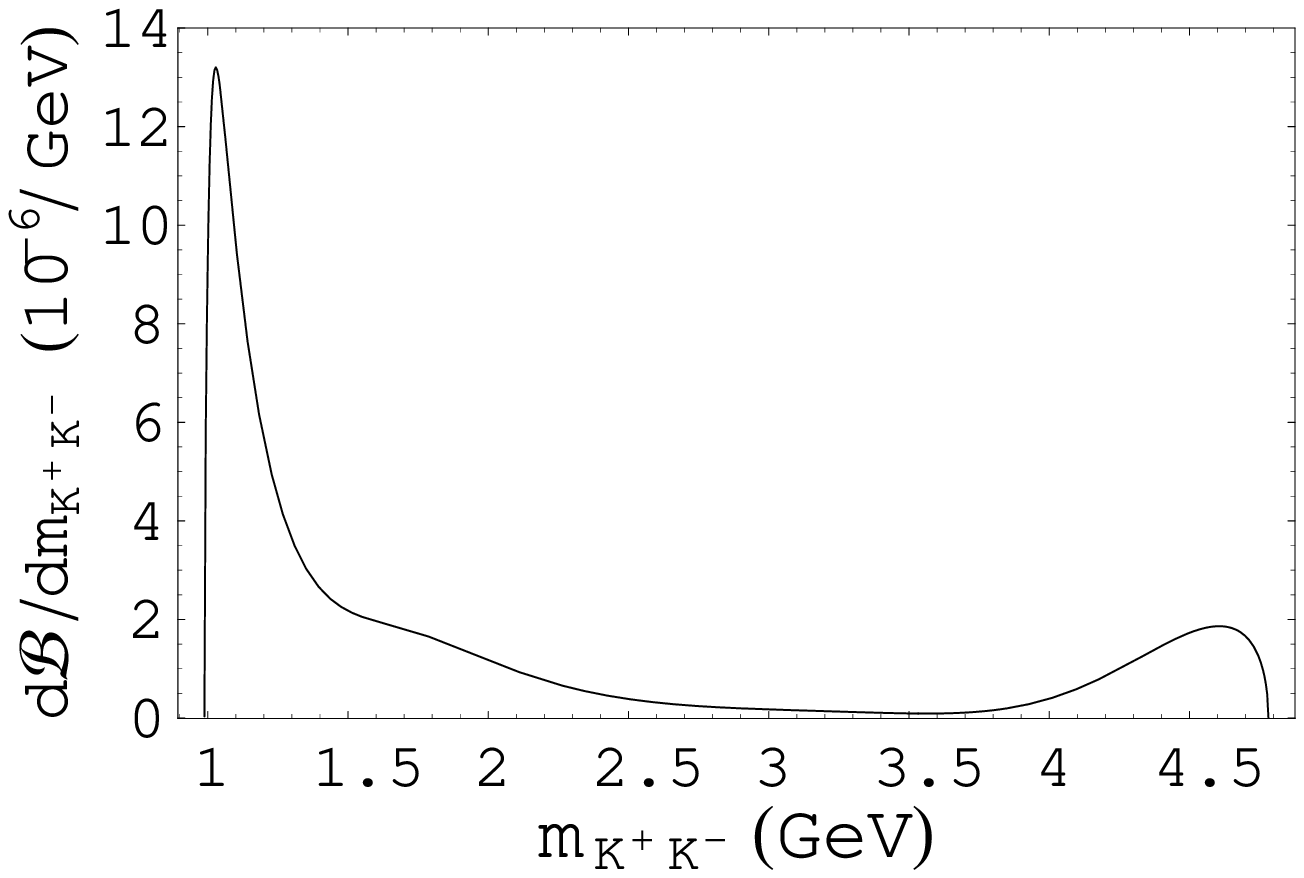,width=8cm}
              \hspace{0.2cm}
              \epsfig{figure=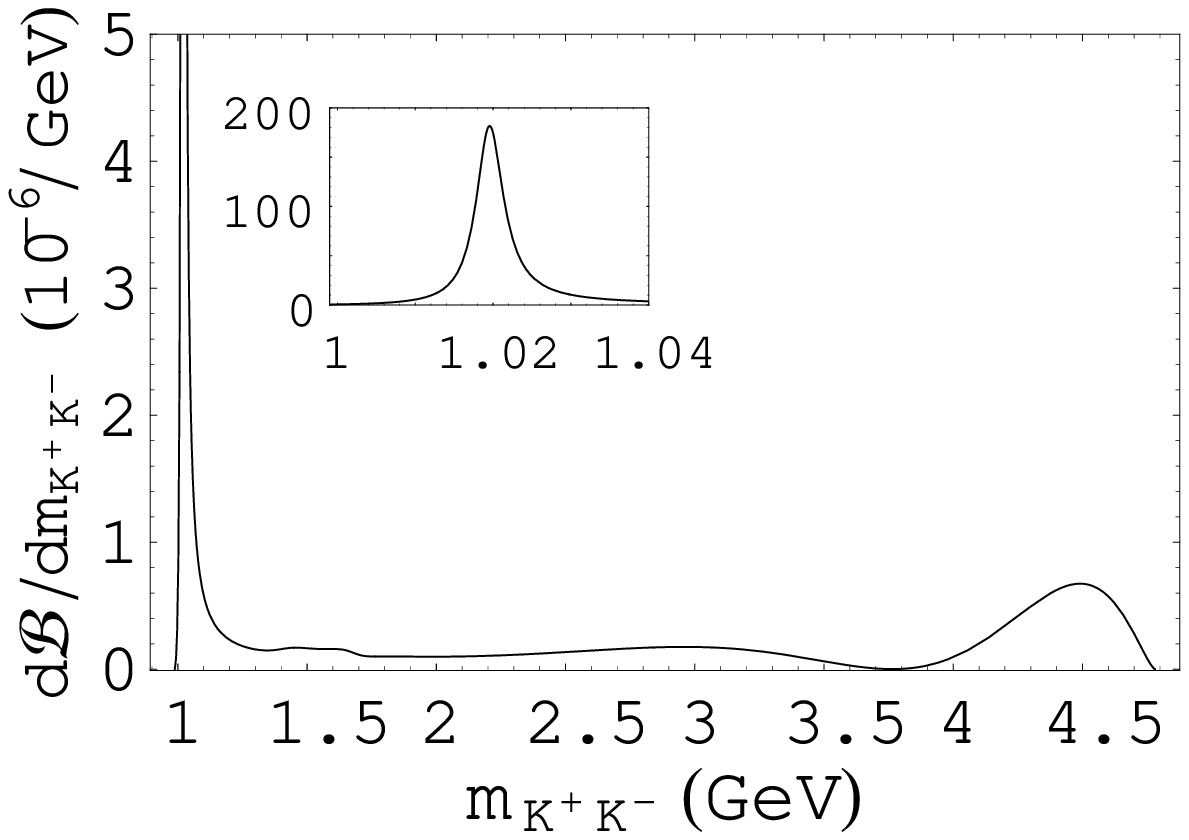,width=8.2cm}
              }
  \centerline{(a)
              \hspace{8cm}
              (b)
              }
    \caption{\small The $K^+ K^-$ mass spectra for
    $\overline B {}^0\to K^+ K^- K_S$ decay from (a) $CP$-even and (b) $CP$-odd contributions.
    The insert in (b) is for the $\phi$ region. Results for
$(K^+K^-K_L)_{CP\pm}$ are identical to those for
$(K^+K^-K_S)_{CP\mp}$.}\label{fig:rates}
\end{figure}

The $K^+K^-$ mass spectra of the $\overline B {}^0\to K^+ K^- K_S$
decay from $CP$-even and $CP$-odd contributions are shown in
Fig.~\ref{fig:rates}. In the spectra, there are peaks at the
threshold and a milder one in the large $m_{K^+K^-}$ region. For
the $CP$-even part, the threshold enhancement arises from the
$f_0(980) K_S$ and the nonresonant $f_S^{K^+K^-}$ contributions
[see Eq. (\ref{eq:KKsff})], while the peak at large $m_{K^+K^-}$
comes from the nonresonant two-meson transition $\overline B
{}^0\to K^+ K_S$ followed by a current produced $K^-$. Since the
nonresonant term [Eq.~(\ref{eq:omegaNR})] favors a small $m_{K^+
K_s}$ region, the spectrum should peak at the large $m_{K^+ K^-}$
end. For the $CP$-odd spectrum the bump at the large $m_{K^+ K^-}$
end originates from the same two-meson transition term, while the
peak on the lower end corresponds to the $\phi K_s$ contribution,
which is also shown in the insert. The full $K^+K^-K_S$ spectrum
is basically the sum of the $CP$-even and the $CP$-odd parts. Note
that although we include $f_0(1530) K_S$ contribution, its effect
is not as prominent as one may expect from the $K^-K^+K^-$
spectrum where a large $f_X(1500) K^-$ contribution is
found~\cite{Garmash:2004wa}.

\begin{figure}[t]
  \centerline{\epsfig{figure=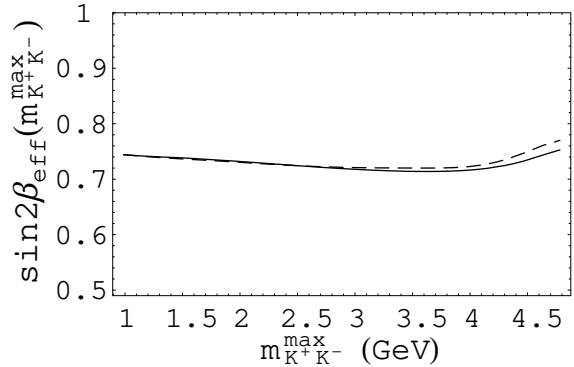,width=8cm}}
      \caption{\small Mixing-induced \CP asymmetry $\sin2\beta_{\rm
eff}(m_{K^+K^-}^{\rm max})$ (see the text for the definition)
versus the invariant mass $m^{\rm max}_{K^+K^-}$ for $K^+ K^- K_S$
with $\phi K_S$ excluded (solid line) and for \CPP-even $K^+ K^-
K_S$ (dashed line). When $m_{K^+K^-}^{\rm max}$ approaches the
upper limit $m_B-m_{K_S}$, the whole phase space is saturated and
$\sin2\beta_{\rm eff}(m_{K^+K^-}^{\rm max})$ is reduced to the
usual $\sin2\beta_{\rm eff}$. This result also applies to the
$K^+K^-K_L$ mode.}
      \label{fig:sin2beta}
\end{figure}

For the mixing-induced \CP asymmetry in the $K^+K^-K_S$ mode, we
compute the effective $\sin 2\beta$ in two different ways: In one
way, we calculate $S$ with $\phi K_S$ excluded in $K^+K^-K_S$ and
then apply the relation $S=-(2f_+-1)\sin 2\beta_{\rm eff}$ and the
theoretical value of $f_+$ to obtain $\sin 2\beta_{\rm eff}$. This
procedure follows closely the BaBar and Belle method of measuring
the effective $\sin 2\beta$. In the other way, we calculate $S$
directly for the \CPP-even $K^+K^-K_S$ and identify $S_{KKK,CP+}$
with $\sin 2\beta_{\rm eff}$. As for the $K_SK_SK_S$ mode, there
is no such ambiguity as it is a purely \CPP-even state. As shown
in Table \ref{tab:AS} and Fig. \ref{fig:sin2beta}, the resulting
$\sin 2\beta_{\rm eff}$ is slightly different in these two
different approaches.

The deviation of the mixing-induced \CP asymmetry in $B^0\to
K^+K^-K_S$ and $K_SK_SK_S$ from that measured in $B\to J/\psi K_S$
(or the fitted CKM's $\sin2\beta$~\cite{CKMfitter}), namely,
$\Delta \sin 2\beta_{\rm eff}\equiv \sin 2\beta_{\rm eff}-\sin 2
\beta_{J/\psi K_S\,(CKM)}$, is calculated from Table \ref{tab:AS}
to be
 \be
 \label{eq:DeltaS}
 \Delta \sin 2\beta_{K^+K^-K_S}=0.06^{+0.08}_{-0.02}\,\,(0.02^{+0.08}_{-0.02}),\qquad
 \Delta \sin 2\beta_{K_SK_SK_S}=0.06^{+0.00}_{-0.00}\,\,(0.02^{+0.00}_{-0.00}).
 \en
Note that part of the deviation comes from that between the
measured $\sin 2 \beta_{J/\psi K_S}$ and the fitted CKM's
$\sin2\beta$. The $K^+K^-K_S$ has a potentially sizable
$\Delta\sin 2\beta$, as this penguin-dominated mode is subject to
a tree pollution due to the presence of color-allowed tree
contributions. For the $K_SK_SK_S$ mode, the central value and the
error on $\Delta\sin 2\beta$ are small.

It is instructive to see the dependence of $\sin 2\beta_{\rm eff}$
on the $K^+K^-$ invariant mass, $m_{K^+K^-}\equiv
m_{23}=\sqrt{s_{23}}$. For the phase space integration in Eq.
(\ref{eq:A&S}), for a given $s_{23}$, the upper and lower bounds
of $s_{12}$ are fixed. The invariant mass $m_{23}$ is integrated
from $m_{23}^-=m_2+m_3$ to $m_{23}^+=m_B-m_1$. When the variable
$s_{23}$ or $m_{23}$ is integrated from $m_{23}^-$ to a fixed
$m_{23}^{\rm max}$ (of course, $m_{23}^-< m_{23}^{\rm max}\leq
m_{23}^+$), the effective $\sin 2 \beta$ thus obtained is
designated as $\sin 2\beta_{\rm eff}(m_{23}^{\rm max})$. Fig.
\ref{fig:sin2beta} shows the plot of $\sin 2 \beta_{\rm
eff}(m_{K^+K^-}^{\rm max})$ versus $m^{\rm max}_{K^+K^-}$ for
$K^+K^-K_S$. Since there are two different methods for the
determination of $\sin 2\beta_{\rm eff}$, the results are depicted
in two different curves. It is interesting that $\sin
2\beta(m_{23}^{\rm max})$ is slightly below $\sin 2\beta_{CKM}$ at
the bulk of the $m_{K^+K^-}$ region and gradually increases and
becomes slightly larger than $\sin 2\beta_{CKM}$ when the phase
space is getting saturated. The deviation $\Delta
\sin2\beta_{K^+K^-K_S}$ arises mainly from the large $m_{K^+K^-}$
region.

Direct \CP violation is found to be very small in both $K^+K^-K_S$
and $K_SK_SK_S$ modes. It is interesting to notice that direct \CP
asymmetry in the \CPP-even $K^+K^-K_S$ mode is only of order
$10^{-3}$, but it becomes $0.2\times 10^{-2}$ in $K^+K^-K_S$ with
$\phi K_S$ excluded. Since these direct \CP asymmetries are so
small they can be used as approximate null tests of the SM.

Since direct \CP violation in charmless $B$ decays can be
significantly affected by final-state rescattering \cite{CCS}, we
have studied to what extent indications of possibly large
deviations of the mixing-induced \CP violation seen in the
penguin-induced two-body decay modes from $\sin 2\beta$ determined
from $B\to J/\psi K_S$ can be accounted for by final-state
interactions \cite{CCSsin2beta}. It is natural to extend the study
of final-state rescattering effect on time-dependent \CP
asymmetries to $B\to KKK_S$ decays. Final-state interactions in
three-body decays are expected to be much more complicated than
the two-body case.
For example, the color allowed tree decay $\ov B {}^0\to
D^{(*)+}_s D^{(*)-}$ can rescatter into a $K^+ K^- K_S$ final
state, where we have $D^{(*)+}_s\to K^+ \bar D^{*0}$, $D^{(*)-}\to
K_S D^{*-}_s$ followed by a $\bar D^{*0} D^*_s\to K^-$ fusion.
These diagrams are too complicated and will not be included in
this study.\footnote{In passing we note that these diagrams could
have the effect of increasing somewhat our predictions for the
rates of 3$K$ final states. Although these contributions carry
negligible \CPP-odd (weak) phases, they also contribute to the
strong phases and hence will tend to dilute our prediction on
$\Delta\S$ but not necessarily on direct \CP asymmetries.}
Nevertheless, we attempt to incorporate final state rescattering
effects in a simple way by including resonance contributions to
the corresponding kaon pairs in the final state \cite{hewett-res}.
We note that another attempt in this direction has recently been
made by Furman {\it et al.} \cite{Furman}. They considered
rescattering of $\pi\pi$ and $K\ov K$ pairs in the $\pi\pi$
effective mass range from threshold to 1.1 GeV. While their
predicted direct \CP asymmetry is very small, the parameter $\S$
is found to be $-0.64$ or $-0.77$, depending on the set of penguin
amplitudes. However, due to the limitation on phase space, the
calculated branching ratios of order $1\times 10^{-6}$ for
$K^+K^-K_S$ and $K_SK_SK_S$ are only small portions of the total
experimental rates (see Table \ref{tab:Data}) and, consequently,
the predictions of $S$ may be affected when the whole phase space
is taken into consideration.

\section{Conclusions}
In the present work we have studied the decay rates and
time-dependent \CP asymmetries in the decays $B^0\to
K^+K^-K_{S(L)}$ and $K_S K_S K_{S(L)}$ within the framework of
factorization. Our main results are as follows:

\begin{enumerate}

\item Resonant and nonresonant contributions to the hadronic
matrix elements are carefully investigated. We incorporate final
state rescattering effects in a simple way by including resonance
contributions to the corresponding kaon pairs in the final state.
Instead of applying heavy meson chiral perturbation theory to the
matrix element for $B\to KK$, which is valid only for a small
kinematic region, we consider the resonant contribution from the
$B_s^*$ pole and nonresonant contributions constrained by QCD
counting rules.

\item Using the $K_SK_SK_S$ decay rate as an input, the predicted
branching ratio of $K^+K^-K_{S(L)}$ modes and the \CPP-even (-odd)
fraction of $B^0\to K^+K^-K_{S(L)}$ are consistent with the data
within the theoretical and experimental errors, though the
theoretical central values on rates are somewhat smaller than the
experimental ones.

\item Owing to the presence of color-allowed tree contributions in
$B^0\to K^+K^-K_{S(L)}$, this penguin-dominated mode is subject to
a potentially significant tree pollution and the deviation of the
mixing-induced \CP asymmetry from that measured in $B\to J/\psi
K_S$, namely, $\Delta \sin 2\beta_{K^+K^-K_{S(L)}}\equiv \sin
2\beta_{K^+K^-K_{S(L)}}-\sin 2 \beta_{J/\psi K_S}$, can be as
large as ${\cal O}(0.10)$. The deviation $\Delta
\sin2\beta_{K^+K^-K_{S(L)}}$ arises mainly from the large
$m_{K^+K^-}$ region.

\item The $K_S K_S K_{S(L)}$ mode appears theoretically very clean
in our picture: The uncertainties in $\Delta \sin 2\beta_{\rm
eff}$ are negligible.

\item Direct \CP asymmetries are very small in both
$K^+K^-K_{S(L)}$ and $K_SK_SK_{S(L)}$ modes.

\end{enumerate}

\vskip 2.0cm \acknowledgments  We wish to thank Kai-Feng Chen for
discussion. This research was supported in part by the National
Science Council of R.O.C. under Grant Nos. NSC93-2112-M-001-043,
NSC93-2112-M-001-053 and by the U.S. DOE contract No.
DE-AC02-98CH10886(BNL).

\vskip 2.0cm {\it Note added:} After the paper was submitted for
publication, BaBar has presented a Dalitz plot study of $B^0\to
K^+K^-K_S^0$ decays \cite{BaBarKKKsEPS}. The BaBar results
constrain the tree contribution (incorporated via Eq. (2.18) in
the present work) in rates and, as a result, a small $\Delta\sin
2\beta_{K^+K^-K_S}$ is preferable.


\end{document}